\documentclass[9pt,twocolumn,twoside]{osajnl}

\journal{ol}

\setboolean{shortarticle}{true} 

\usepackage{amsmath,amssymb}
\usepackage{graphicx}
\usepackage{color}
\usepackage{bm}

\newcommand{\dif}{\mathrm{d}}

\def\myrev#1{#1}
\def\mydel#1{}

\title{Observation of microscale nonparaxial optical bottle beams}

\author[1]{Raluca-Sorina Penciu}
\author[2]{Yujie Qiu}
\author[1]{Michael Goutsoulas}
\author[2]{Xiaopei Sun}
\author[2]{Yi Hu}
\author[2]{Jingjun Xu}
\author[2,3]{Zhigang Chen}
\author[1,4,*]{Nikolaos K. Efremidis}

\affil[1]{Department of Mathematics and Applied Mathematics, University of Crete, 70013 Heraklion, Crete, Greece}
\affil[2]{The  MOE  Key  Laboratory  of  Weak-Light  Nonlinear  Photonics,  and  TEDA  Applied  Physics  Institute  and  School  of Physics, Nankai University, Tianjin 300457, China}
\affil[3]{Department of Physics and Astronomy, San Francisco State University, San Francisco, CA 94132}
\affil[4]{Institute of Applied and Computational Mathematics, Foundation for Research and Technology - Hellas (FORTH), 70013 Heraklion, Crete, Greece.}
\affil[*]{nefrem@uoc.gr}

\ociscodes{
(050.1940) Diffraction;
(070.7345) Wave propagation; 
(260.2110) Electromagnetic optics; 
(350.7420) Waves.}

\begin{abstract}
We predict and experimentally observe three-dimensional microscale nonparaxial optical bottle beams based on the generation of a caustic surface under revolution. Such bottle beams exhibit high contrast between the surrounding surface and the effectively void interior. Via caustic engineering we can precisely control the functional form of the high intensity surface to achieve microscale bottle beams with longitudinal and transverse dimensions of the same order of magnitude. 
Although, in general, the phase profile at the input plane can be computed numerically, we find closed form expressions for bottle beams with various type of surfaces both in real and in the Fourier space. 
\end{abstract}

\setboolean{displaycopyright}{true}

\begin{document}

\maketitle

An optical bottle beam has a low intensity central spot that is surrounded by a high intensity region in three-dimensional space. Since its prediction~\cite{arlt-ol2000} several techniques have been proposed for the generation of such beams~\cite{rudy-oe2001,mcglo-oc2003,ahluw-oc2004,ahluw-oe2004,yelin-ol2004,pu-oc2005,shved-oe2009,zhang-ol2011-bottle,alpma-apl2012}. These methods mainly utilize on-axis destructive interference of beams at the focus with different parameters/properties. Optical bottle beams are mainly used in particle manipulation for trapping and transporting particles. An alternative method for the generation of bottle beams that relies on the use of caustics under revolution has been suggested and implemented in~\cite{chrem-ol2011-fourier,chrem-pra2012}. Such bottle beams are closely related to the effect of abrupt autofocusing~\cite{efrem-ol2010,papaz-ol2011,zhang-ol2011} of accelerating beams~\cite{sivil-ol2007,sivil-prl2007}. A main advantage in this case is that the radius of the surrounding high-intensity surface can be designed to follow arbitrary convex shapes with precision as a function of the propagation distance ($r(z)$). However, such bottle beams are thus far generated in the paraxial regime and, as a result, they have an elongated profile not favorable for single-particle trapping and manipulation.

In this Letter we predict and experimentally observe nonparaxial optical bottle beams at the microscale that are generated by a caustic surface under revolution. The resulting waves have high intensity contrast between the void region and the surrounding surface, and bottle shape that can be engineered with precision even at the microscale. In addition, the length scales in the transverse and longitudinal directions can be made to be comparable. In general, the phase profile at the input plane is obtained numerically. Here we find closed-form expressions for bottle beams with spherical, ellipsoid of revolution, and paraboloid surfaces that are generated either in the real or in the Fourier space.

We start by considering the beam propagation of an optical wave in a linear, homogeneous, and isotropic medium and follow a similar procedure as in~\cite{penci-ol2016}. Utilizing Gauss' law $\nabla \cdot {\bm D} =0 $, the electric field is expressed as
\begin{align}
{\bm E}=-\frac{1}{\epsilon} \nabla\times {\bm F}, \label{eq:E0}
\end{align}
where ${\bm F}$ is an auxiliary vector potential and $\epsilon$ the electric permittivity. In the case of monochromatic waves and assuming a linear polarization for the auxiliary field $\bm F = \hat{\bm y}F$, we find that $F$ satisfies the Helmholtz equation 
\begin{equation}
(\nabla^2+k^2) F = 0 
\label{eq:Helmholtz},
\end{equation}
where $k=n\omega/c=2 \pi/\lambda$, $\nabla^2=\partial_x^2+\partial_y^2+\partial_z^2$, $(x,y)$ and $(r,\theta)$ are the transverse coordinates in Cartesian and polar form, and $z$ is the propagation coordinate. We assume that at the input plane $\bm F$ is radially symmetric \myrev{$\bm F(z=0)=\hat{\bm y} F(r,z=0)$}. Due to the symmetries of the Helmholtz equation the auxiliary vector potential maintains its radial profile upon propagation, i.e., $\bm F = \hat{\bm y}F(r,z)$. Substitution to Eq.~(\ref{eq:E0}) then leads to 
\begin{align}
{\bm E}=(1/\epsilon)(\hat{{\bm x}} \partial_z F
-\hat{{\bm z}} \cos\theta \partial_r F).
\label{eq:E}
\end{align}
In Eq.~(\ref{eq:E}) we notice that the $x$ component of the electric field is radially symmetric, whereas the $z$ component exhibits a dipolar structure. In the rest of the paper we prefer to depict the intensity along the $x-z$ plane where the amplitude attains a maximum as a function of $\theta$. 

The solution of Eq.~(\ref{eq:Helmholtz}) is given by the following Hankel transform pair
\begin{equation}
F(r,z)=\frac1{2\pi}
\int_0^\infty
\tilde F(k_\perp)
J_0(k_\perp r)
e^{ik_zz}
k_\perp\,\dif k_\perp,
\label{eq:H1}
\end{equation}
\begin{equation}
\tilde F(k_\perp)=2\pi
\int_0^\infty
F(\rho,z=0)
J_0(k_\perp\rho)
\rho\,\dif \rho
\label{eq:H2}
\end{equation}
\myrev{with $\rho$ being the radial polar coordinate at the input plane ($\rho=r(z=0)$).}
Using large argument asymptotics for the Bessel function, decomposing the fields in Eqs.~(\ref{eq:H1})-(\ref{eq:H2}) into amplitude and phase as $F(\rho,z=0)=A(\rho)e^{i\phi(\rho)}$, $\tilde F(k_x)=\tilde A(k_x)e^{i\tilde\phi(k_x)}$, and applying first and second order stationarity of the phase we derive the ray equation
\begin{equation}
r=\rho+\frac{\phi'(\rho)z}{k_z(\phi'(\rho))}
\label{eq:ray}
\end{equation}
and the relation for the high intensity surface of the beam under revolution. More importantly, we can find the phase at the input plane
\begin{equation}
\phi'(\rho)=
\frac{k(\dif f/\dif z_c)}{(1+(\dif f/\dif z_c)^2)^{1/2}}
\label{eq:tangent}
\end{equation}
that is required for a beam that forms the predefined caustic surface under revolution ($z$ is the axis of rotation) $r_c=f(z_c)$ where $\rho = f(z_c)-z_cf'(z_c)$, and the subscript $c$ stands for caustic. In general Eq.~(\ref{eq:tangent}) can be integrated numerically. Below we will consider three different classes of caustic surfaces, namely spherical, ellipsoid of revolution, and paraboloid that are associated with closed-form expressions for the phase at the input plane. 

\begin{figure}
\centerline{\includegraphics[width=\columnwidth]{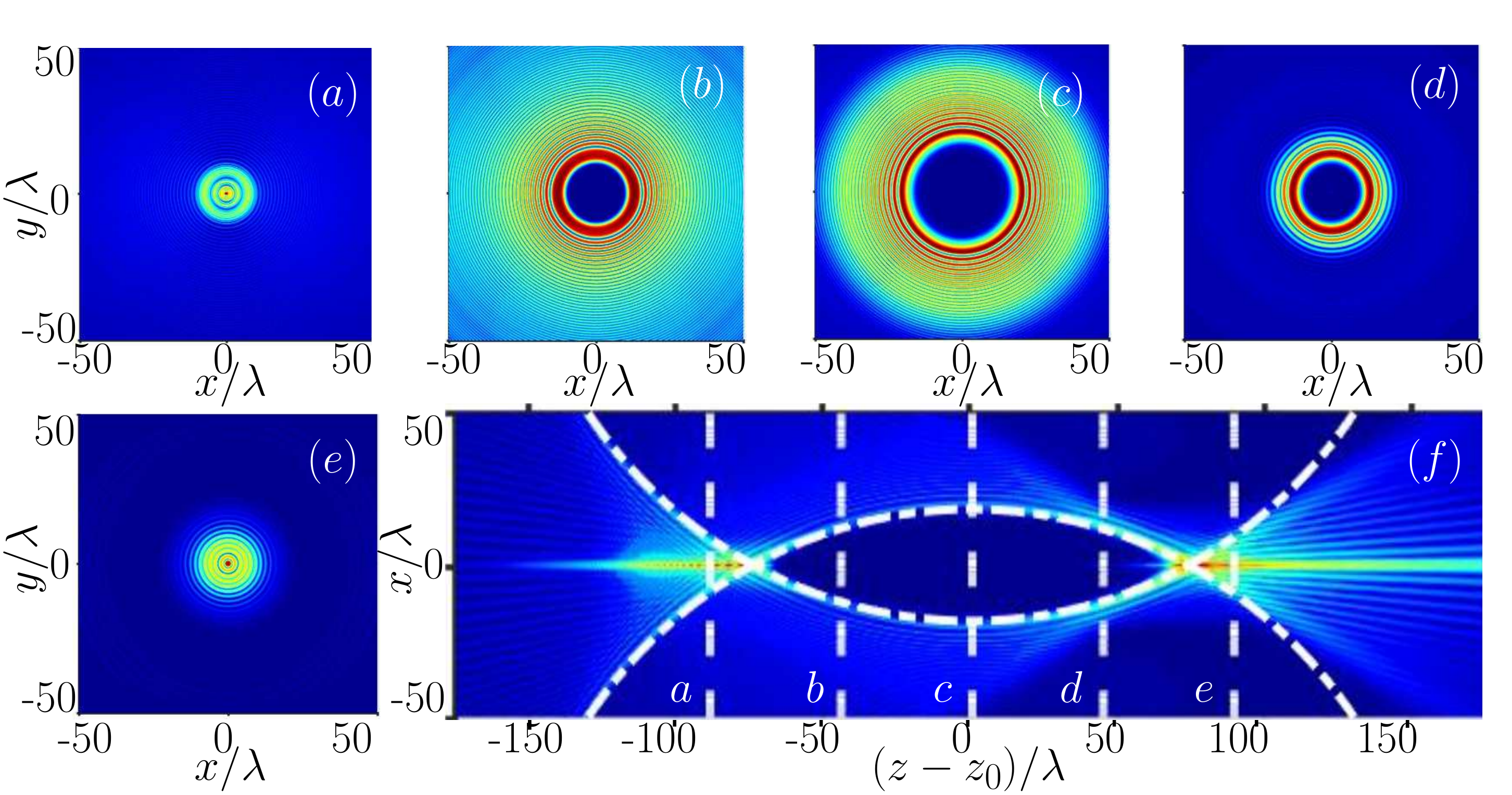}}
\caption{An optical bottle beam with a spherical surface having $R=158\lambda$, $r_0=140\lambda$ and $z_0=175\lambda$. In (f) the intensity distribution is shown in the $y=0$ plane in logarithmic scale. The white dash-dotted curve is the theoretical design for the surface of the bottle and the vertical white dashed lines denote the different intensity cross sections in the $x-y$ plane shown in (a)-(e).  \label{fig:bottle_circ_theory}}
\end{figure}
We start with the case of a bottle beam with spherical surface $r(z) = (R^2 - (z-z_0)^2)^{1/2} - r_0$ where $R$ is the radius, and $(-r_0,z_0)$ is the center (and thus $R>r_0$). The bottle beam extends in the longitudinal direction between $z_0\pm(R^2-r_0^2)^{1/2}$, its length is $L_z=2(R^2-r_0^2)^{1/2}$, while its diameter is $D=2(R-r_0)$. The condition $z_0>(R^2-r_0^2)^{1/2}$ should be satisfied so that the entire bottle is located after the plane of incidence. Following the relevant calculations, from Eq.~(\ref{eq:tangent}) we obtain 
\begin{equation}
F(\rho,z=0) = A(\rho) \left( e^{i \Phi(r_0+\rho)} + e^{i \Phi(r_0-\rho)} \right),
\end{equation}
where
\begin{equation}
\Phi (x) =  k \left[ R \left( \arctan \left( \frac{x}{z_0} \right)+ \arctan \left( \frac{u(x)}{R} \right) \right)-u(x) \right], 
\label{eq:phi1}
\end{equation}
and $u(x) = (x^2 + z_0^2 -R^2)^{1/2}$.

Typical simulation results are shown in Fig.~\ref{fig:bottle_circ_theory}. 
The initial amplitude distribution is $A(\rho) = 1/\sqrt{\rho + r_0}$ while the aperture $\rho_a$ is selected to be slightly larger than the radius of the ray that is tangent at the second focal point of the bottle. We see the formation of a bottle beam with a spherical surface. The agreement between theory and simulations is very good. In Fig.~\ref{fig:bottle_circ_theory}(a)-(e) the high intensity ring profile at different cross-sections in the transverse plane is depicted. In all of these plots the contrast between the almost void interior of the bottle and the high intensity ring is very high. From Eq.~(\ref{eq:E}) we see that the intensity of \myrev{each} ring is not uniform due to the $z$-component of the electric field. In particular, the \mydel{total} intensity is modulated in a sinusoidal fashion with the angular coordinate taking maximum along the $x$-axis and its minimum along the $y$-axis. However, these slight intensity variations are mild since the $z$-component of the electric field is smaller as compared to the $x$-component.

\begin{figure}
\centerline{\includegraphics[width=\columnwidth]{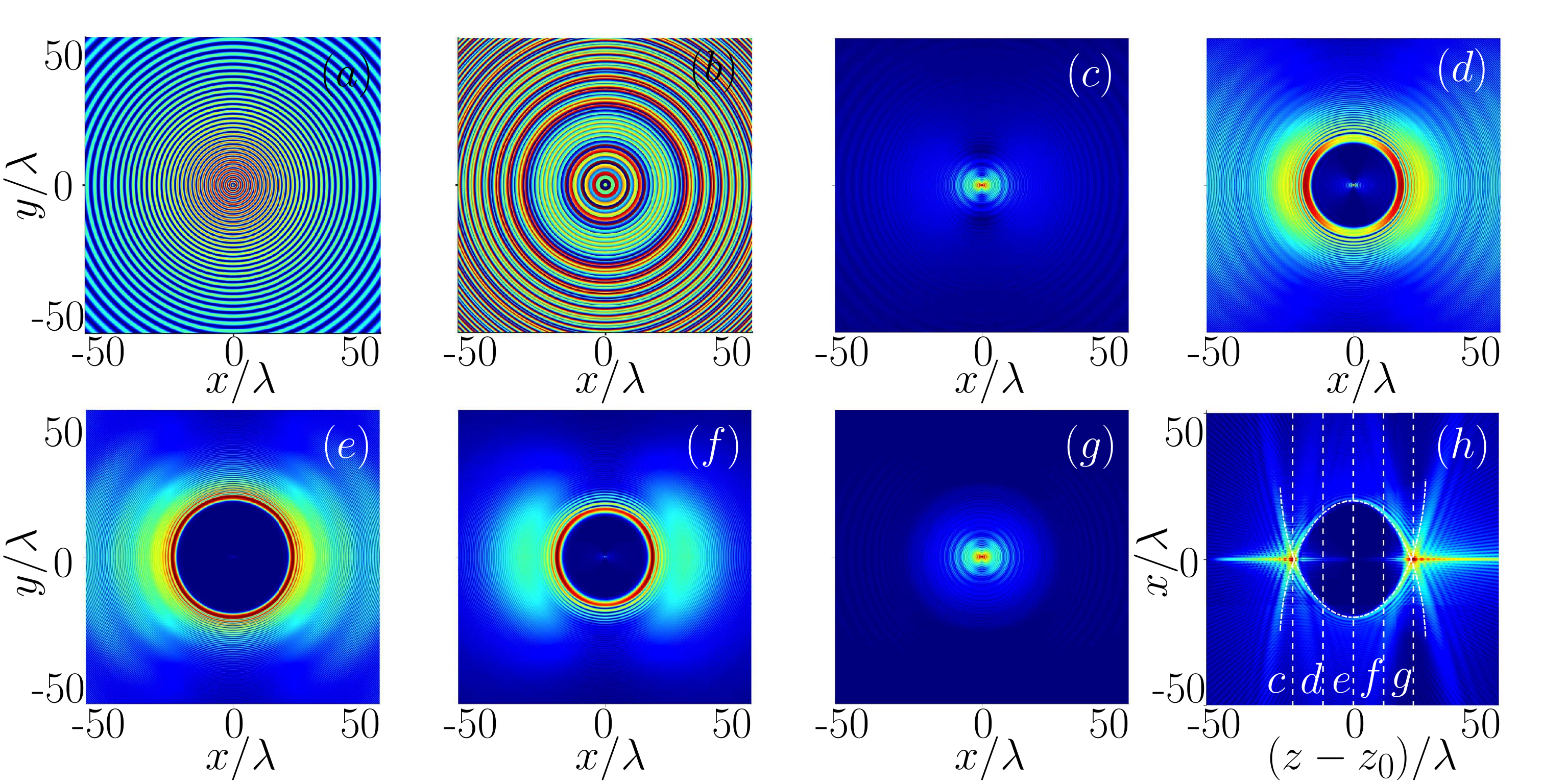}}
\caption{An optical bottle beam with a spheroidal surface having $R=50\lambda$, $r_0=30\lambda$, $z_0=50\lambda$ and $a=0.5$. \myrev{The initial amplitude and phase profile of the auxiliary field $F$ is shown in (a) and (b), respectively.} In (h) the intensity distribution is shown in the $y=0$ plane in logarithmic scale. The white dash-dotted curve is the theoretical design for the surface of the bottle and the vertical white dashed lines denote the different intensity cross sections in the $x-y$ plane shown in (c)-(g). \label{fig:bottle_ellip_theory}}
\end{figure}
Closed form expressions can also be found for bottle beams with a spheroid (ellipsoid of revolution) surface profile. Specifically the radius of the bottle beam is given by the expression
\begin{equation}
r(z) = \sqrt{R^2 - \left( \frac{z-z_0}{a} \right)^2}-r_0.
\end{equation}
In the above equation $(-r_0,z_0)$ is the center and $R$, $a R$ are the semi-axes in the transverse and the longitudinal directions of the spheroid. The bottle extends in the longitudinal direction between $z_0\pm a(R^2-r_0^2)^{1/2}$ and thus its length is $L_z=2a(R^2-r_0^2)^{1/2}$. In the transverse direction its diameter is $D=2(R-r_0)$. Following the relevant calculations we find that the vector potential at the input plane should be given by 
\begin{equation}
F(\rho,z=0) = A(\rho) \left( e^{i \Phi(\sigma_-(r_0+\rho))} + e^{i \Phi(\sigma_+(r_0-\rho))} \right),
\end{equation}
where the associated phase is 
\begin{multline}
\Phi (\sigma) = \frac{k a}{\sigma^2-1} \left( \sqrt{1-\sigma^2} \sqrt{1+\left( \frac{1}{a^2}-1 \right)\sigma^2 } \left(\frac{z_0}{a}-R \sigma \right) + \right. \\ 
 \left.(1-\sigma^2) R E \left[ \sin^{-1}( \sigma) \left| 1-\frac{1}{a^2} \right. \right]  \right),
\end{multline}
\begin{equation}
\sigma_\pm (\rho) = \frac{R \frac{z_0}{a} \pm \rho \sqrt{\rho^2+\left( \frac{z_0}{a} \right)^2 -R^2}} {\rho^2+\left( \frac{z_0}{a} \right)^2}, 
\label{eq:sigma_pm}
\end{equation}
and $E[\phi|m]$ is the elliptic integral. Simulation results in the case of an oblate ($a<1$) spheroidal bottle beam are shown in Fig.~\ref{fig:bottle_ellip_theory}. The initial amplitude profile is selected as $A(\rho) = 1/\sqrt{\rho + r_0}$. This is a strongly nonparaxial bottle beam with $s=L_z/D=1$. In comparison to the spherical case, here the $z$-component of the electric field is stronger leading to a visible sinusoidal modulation of the intensity in the transverse plane as a function of the angular coordinate. 

\begin{figure}
\centerline{\includegraphics[width=\linewidth]{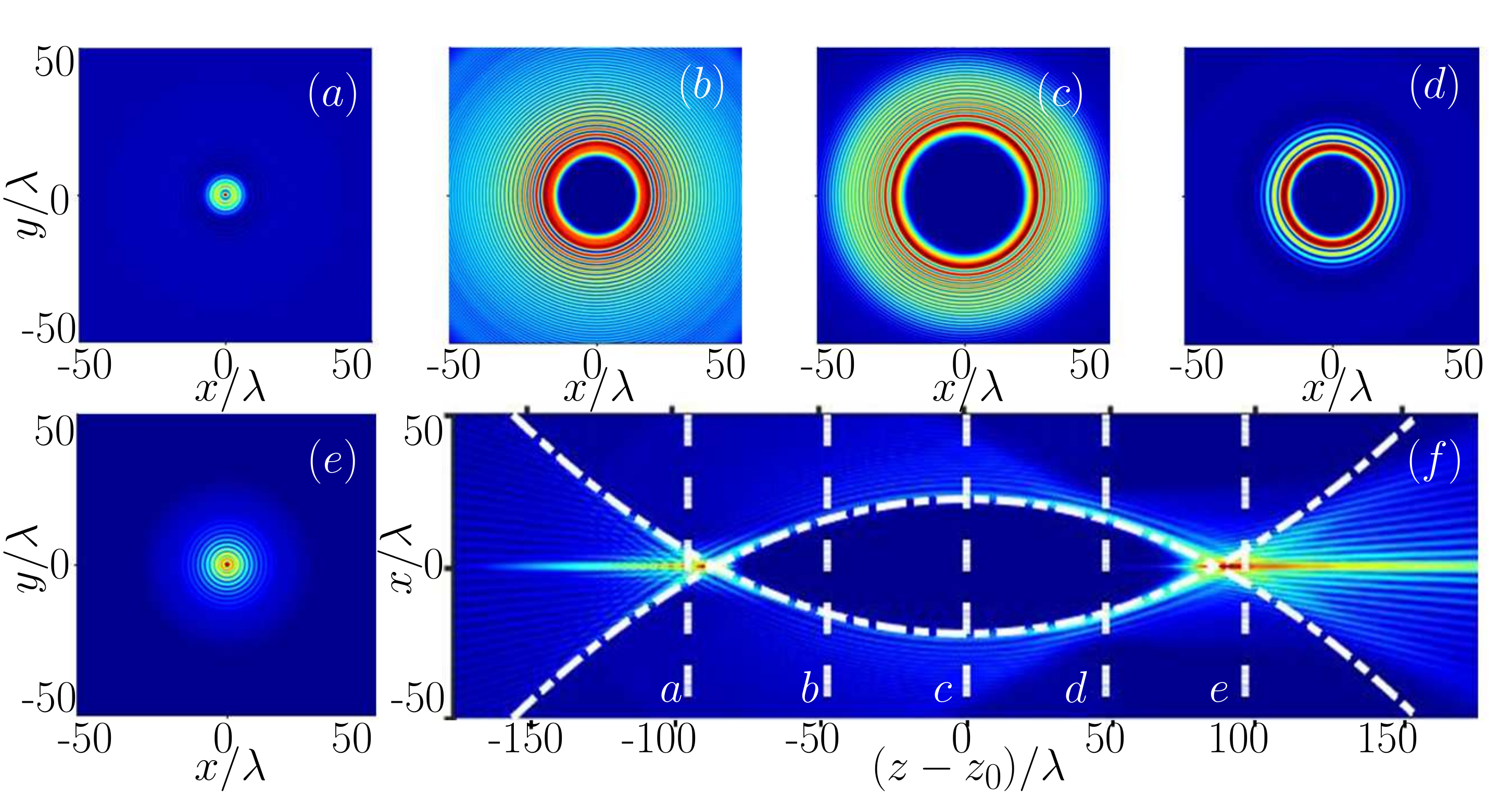}}
\caption{An optical bottle beam with a paraboloidal surface having $c=69.7\lambda$, $a=0.003/\lambda$, $b=1.05$, $z_0=175\lambda$. In (f) the intensity distribution is shown in the $y=0$ plane in logarithmic scale. The white dash-dotted curve is the theoretical design for the surface of the bottle and the vertical white dashed lines denotes the different intensity cross sections in the $x-y$ plane shown in (a)-(e). \label{fig:bottle_parab_theory}}
\end{figure}
We have also found closed-form expressions for bottle beams with paraboloid surfaces of the form
\begin{equation}
r(z) = -a z^2+b z-c.
\end{equation}
The above equation can be expressed as
\[
r_0-r=(z-z_0)^2/L
\]
where $D=2r_0=2a(b^2/(4a^2)-c/a)$ is the diameter of the bottle beam in the transverse plane at $z_0=b/(2a)$, $L=1/a$ determines the curvature of the surface, and its length is $L_z=2\sqrt{r_0L}$.
Following the relevant calculations we find that the phase of the auxiliary field 
\begin{equation}
F(\rho,z=0) = A(\rho) \left( e^{i \Phi(c+\rho)} + e^{i \Phi(c-\rho)} \right),
\end{equation}
is given by
\begin{equation}
\Phi (x) = - \frac{k}{4 a} \left[  v_+(x) \sqrt{1+ v_-(x)^2} + \sinh^{-1}( v_+(x)) \right],
\end{equation}
with $v_\pm(x) =b \pm 2 \sqrt{a x} $.
The existence condition for the paraboloid bottle-beam $r_0>0$ translates to $b > \sqrt{4 a c}$. Figure~\ref{fig:bottle_parab_theory} presents simulation results of a paraboloid bottle beam with a ratio $s=L_z/D=3.87$. The initial field amplitude is $A(\rho) = 1/\sqrt{\rho + c}$.

The generation of bottle beams as described above requires the modulation of both the amplitude and the phase of the beam at the input plane. However, as shown in~\cite{chrem-ol2011-fourier,chrem-pra2012}, the experimental complexity is significantly reduced if such beams are generated in the Fourier space. Specifically, it is sufficient to utilize the natural Gaussian mode of the laser and modulate only the phase. The following equation relates the phase in the Fourier space with the trajectory of the beam~\cite{mathi-ol2013}
\begin{equation}
\frac{\partial \Phi(\rho)}{\partial \rho}= \frac{r(z)-z \partial_z r(z)}{f}
\label{eq:fs1}
\end{equation}
where
\begin{equation}
\rho = -f \frac{\partial_z r(z)}{\sqrt{1+[\partial_z r(z)]^2}},
\label{eq:fs2}
\end{equation}
and $f$ is the focal length of the objective. In our calculations we select the following initial condition in the Fourier space
\begin{equation}
F(\rho,z=0) =  A(\rho)[e^{-i \Phi(\rho)} + e^{-i \Phi(-\rho)}].
\end{equation}
In the three classes of nonparaxial bottle beams discussed above the Fourier space phase profile can be analytically derived by utilizing Eqs.~(\ref{eq:fs1})-(\ref{eq:fs2}). Specifically, in the case of a bottle with spheroid or spherical ($a=1$) surface the phase is given by
\begin{multline}
\Phi(\rho) = k\{R E[ \sin^{-1}(\xi) | 1 - a^2] -(1-\xi^2)^{1/2} z_0- r_0 \xi \}. 
\end{multline}
where $\xi=\rho/f$. 
Finally the phase for a paraboloidal surface in the Fourier space is
\begin{equation}
\Phi(\rho)=\frac{k}{4a}[(b^2-1) \xi+2b(1-\xi^2)^{1/2}+\tanh^{-1}(\xi)]-kR\xi. 
\end{equation}
where $\xi=\rho/f$.

\begin{figure}
\centerline{\includegraphics[width=\linewidth]{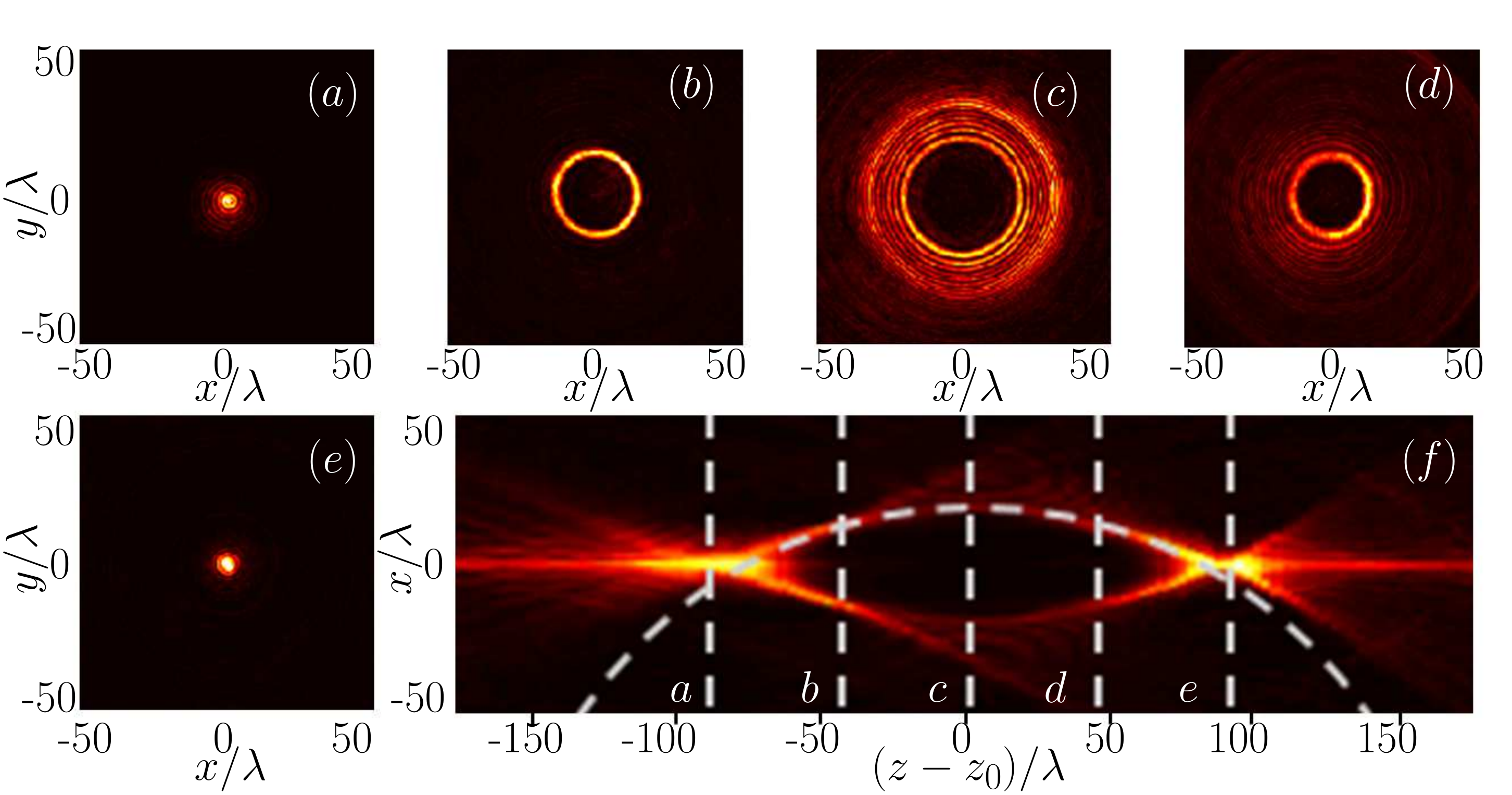}}
\caption{Experimental observation of a spheroidal bottle beam with $\lambda=632.8$ nm and the same parameters as those of Fig.~\ref{fig:bottle_circ_theory}. In (f) the intensity distribution is shown in the $y=0$ plane.
The white dashed curve is the theoretical design for the surface of the bottle and the vertical white dashed lines denote the different intensity cross sections in the $x-y$ plane shown in (a)-(e).}
\end{figure}
\begin{figure}
\centerline{\includegraphics[width=\linewidth]{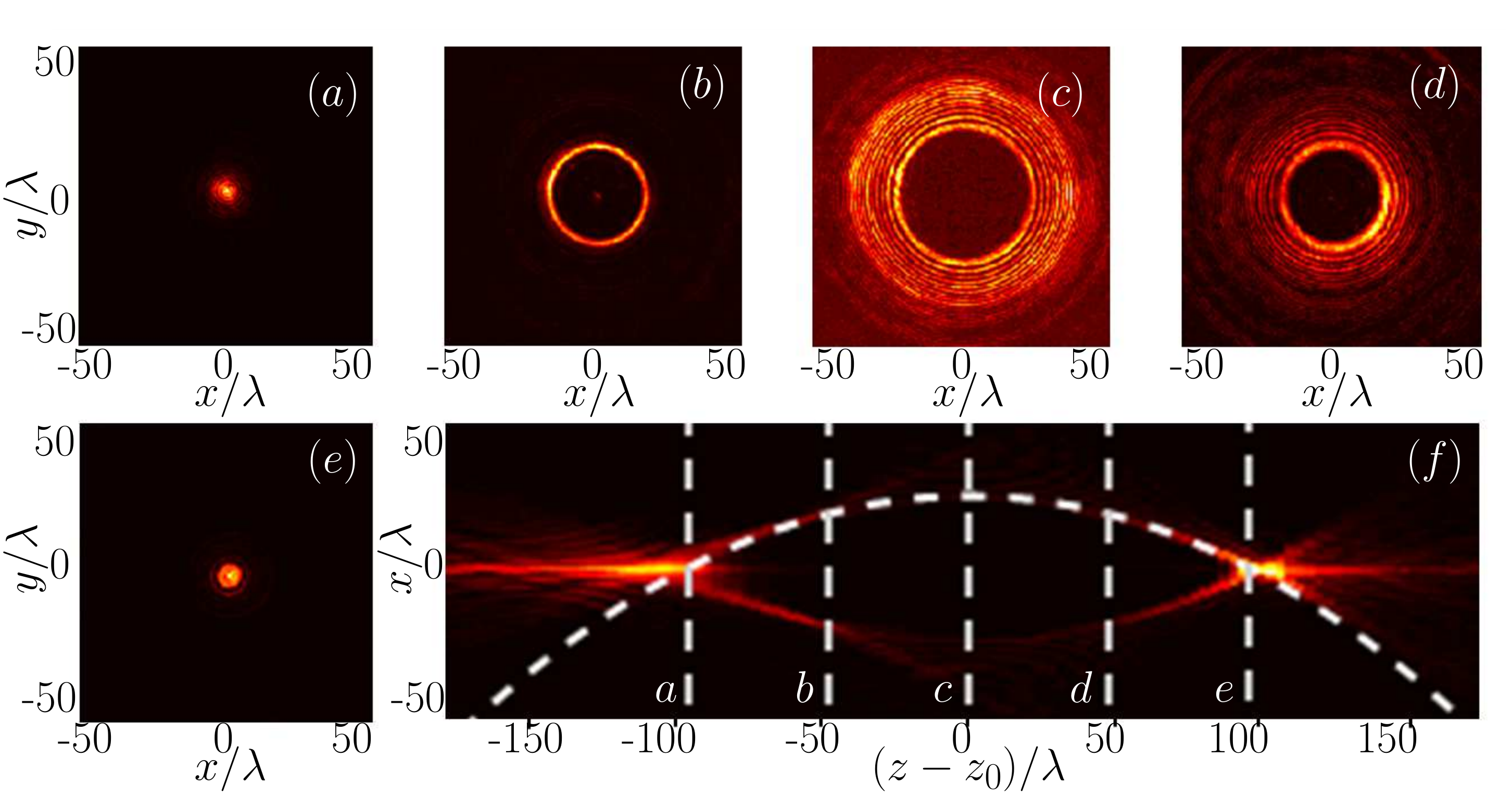}}
\caption{Experimental observation of a paraboloidal bottle beam with $\lambda=632.8$ nm and the same parameters as those of Fig.~\ref{fig:bottle_parab_theory}. In (f) the intensity distribution is shown in the $y=0$ plane.
The white dashed curve is the theoretical design for the surface of the bottle and the vertical white dashed lines denote the different intensity cross sections in the $x-y$ plane shown in (a)-(e). }
\end{figure}

In the experiment, these phases can be wrapped between 0 and 2$\pi$, and generated by employing a computer-assisted spatial light modulator (SLM). Then such a phase modulation is projected into the back focal plane of an objective ($\times60$, NA=0.85). Once a properly aligned broad laser beam ($\lambda=632.8$ nm) illuminates the SLM, the associated bottle beams are produced near the focal plane of the objective. The beam patterns are recorded by using a microscope system (made of another objective and a CCD) positioned on a motorized translation stage. The generated spherical and paraboloidal bottle beams by using the phase calculated from Eqs. (20) and (21) are presented in Fig. 4 and Fig. 5, respectively. Both beams exhibit a high contrast between the void interior and the high intensity surface that follow perfectly along the designed trajectory.
These experimental observations agree well with our theoretical prediction.

In summary, we have predicted and experimentally observed nonparaxial optical bottle beams at the microscale generated by a caustic surface under revolution. Their transverse and longitudinal dimensions can be made to be comparable and thus favorable for single-particle trapping and manipulation. The functional form of the surface 
can be designed to take arbitrary convex shapes while the intensity contrast between the surface and the interior of the bottle is very high.

We acknowledge support
from the Greek State Scholarships Foundation (IKY),
the Erasmus Mundus NANOPHI Project (2013-5659/002-001),
the National Key R\&D Program of China (2017YFA0303800), 
and the NNSF of China (11504186, 61575098, 91750204).

\newcommand{\noopsort[1]}{} \newcommand{\singleletter}[1]{#1}

\end{document}